# Plasmonic monolithic lithium niobate directional coupler switches

*Martin Thomaschewski\*, Vladimir A. Zenin, Christian Wolff, Sergey I. Bozhevolnyi\**

Centre for Nano Optics, University of Southern Denmark, Campusvej 55, DK-5230 Odense M, Denmark

**Abstract.** From the onset of high-speed optical communications, lithium niobite (LN) has been the material of choice for electro-optic modulators owing to its large electro-optic response, wide transparent window, excellent thermal stability and long-term material reliability. Conventional LN electro-optic modulators while continue to be the workhorse of the optoelectronic industry become progressively too bulky, expensive and power hungry to fully serve the needs of this industry rapidly progressing towards highly integrated, cost-effective and energy efficient components and circuits. Recently developed monolithic LN nanophotonic platform enables the realization of electro-optic modulators that are significantly improved in terms of compactness, bandwidth and energy efficiency, while still demanding relatively long, on the mm-scale, interaction lengths. Here we successfully deal with this challenge and demonstrate plasmonic electro-optic directional coupler switches consisting of two closely spaced nm-thin gold nanostripes monolithically fabricated on LN substrates that guide both coupled electromagnetic modes and electrical signals influencing their coupling and thereby enabling ultra-compact switching and modulation functionalities. The extreme confinement of both slow-plasmon modes and electrostatic fields created by two nanostripes along with their nearly perfect spatial overlap allowed us to achieve a 90% modulation depth with 20-$\mu$m-long switches characterized by a electro-optic modulation efficiency of 0.3 V·cm. Our monolithic LN plasmonic platform enables ultra-dense integration of high-performance active photonic components, enabling a wide range of cost-effective optical communication applications demanding $\mu$m-scale footprints, ultrafast operation, robust design and high environmental stability.

## Introduction

In the last decades, lithium niobite has become indispensable for integrated photonics, serving to be the material of choice for electro-optic modulation due to its excellent optical and nonlinear properties. Being advantageous over competing platforms, lithium niobite (LN) fulfills the eligibility material requirements for optical communication systems by exhibiting wide optical transparency (0.35–4.5 μm), large electro-optic coefficients ($r_{33}$ = 30 pm/V) which are preserved at elevated temperatures due to the its high Curie temperature (~1200°C) and excellent chemical and mechanical stability[1]. Leading to considerable commercial significance, the early success of LN for optoelectronic applications was driven by heterogeneous integration of metal diffused channel optical waveguides utilized for chip-scale electro-optic modulators[2–7]. However, the weak confinement of integrated metal-diffused optical waveguides is limiting the electro-optic interaction, resulting in low electro-optic modulation efficiencies and large device footprints. Recently, monolithic integration of thin film lithium niobite modulators[8–14] has attracted an increasing attention due to significant higher optical confinement, leading to improvements in terms of compactness, bandwidth and energy efficiency, while still demanding relatively long, on the mm-scale, interaction lengths due to conceptional limitations in the electro-optic field overlap. Leveraging metal nanostructures to transmit simultaneously both optical and electrical signals, with the additional attribute of extremely enhancing their accompanied local fields by orders of magnitude with excellent overlap, is making plasmonics to a versatile platform for exceptionally compact optoelectronic applications[15,16]. The first pioneering work[17] utilizing surface plasmon polaritons (SPPs) for electrically controlled modulation was based on thermo-optic effects induced by resistive heating in polymer materials. Though this approach facilitates only moderate switching times and relatively high power consumption, the unique property of large overlap between the electromagnetic field of the plasmonic mode and the electrically induced local change of the refractive index was opening the path of exceptionally efficient plasmonic electro-optic modulators. Following this approach, tremendous efforts have been directed towards exploring other electro-optic material platforms which drastically improved the switching performance, including two-dimensional materials[18–21], phase-change materials[22–24] and electro-optic polymers[25–31]. These studies convincingly demonstrate the capability of plasmonics to be a potential complementary technology addressing bottleneck issues in future information technology. However, combining the attractive features of plasmonics with LN[32], to date still the preferred material platform meeting all essential performance requirements, has remained largely unexplored. Here, we demonstrate a monolithic plasmonic modulator based on two identical gold nanostripes on LN, where the metallic structure utilized for applying external electrostatic fields inherently supports the propagation of the surface plasmon polariton (SPP) modes, resulting in an exceptionally simple device architecture. Our approach does not require patterning, etching or milling of the LN substrate, which is challenging in particular due to its mechanical hardness and its chemical stability. The change of the refractive index of LN due to the Pockels electro-optic effect in response to the externally applied electric field on the two gold nanostripes affects the optical coupling between them, which results in a modification of the power distribution between the two ports of a plasmonic directional coupler.

## Results

**Device principles.** The proposed optical switch configuration represents a plasmonic directional coupler that comprises two lithographically fabricated gold nanostripe waveguides placed on z-cut LN, which have identical dimensions of w × h = 350 × 50 nm$^2$ and are separated by the distance of 350 nm (Fig. 1).

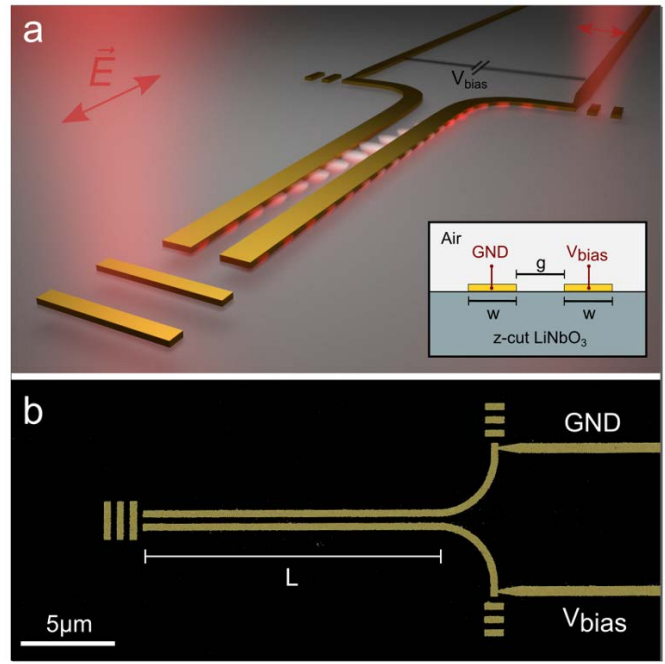

**Figure 1 | Plasmonic monolithic lithium niobite directional coupler switch. a** Conceptual image of power modulation in the plasmonic nanostripes by applying a bias voltage which induces a refractive index change due to the electrooptic effect in the substrate. The inset shows the cross-section of the two identical parallel waveguides of width w = 350 nm, separated by a gap of g = 350 nm placed on z-cut lithium niobate (LiNbO$_3$, LN). **b** Colorized scanning electron microscope (SEM) image of the investigated plasmonic directional coupler switch. The optimal length L of interaction is chosen to be 15.5 µm.

The optically coupled transmission line supports two quasi-TEM modes, i.e., the odd and the even mode, which defines the geometrically dependent coupling length $L_C = 2/[\lambda_0(n_{odd} - n_{even})]$ of the passive system with the mode effective indexes $n_{odd}$ and $n_{even}$, respectively. An optical near-field study is conducted to verify and quantify coupling between the two SPP modes supported by the considered configuration (Supplementary Note 5 and Supplementary Movie 1). In the modulator device, light with the free-space wavelength $\lambda_0$ is fed symmetrically into the two plasmonic waveguides by positioning a diffraction-limited beam on a metallic grating coupler. A 90° bend of the individual waveguides in opposite directions is employed to separate the coupling channels and accordingly define the length $L$ of interaction between the SPP modes supported by nanostripe waveguides. Furthermore, the linearly polarized emission from the terminating coupling gratings is rotated by the bend to be orthogonal to the polarization of the incident beam, thus allowing the cross-polarized far-field imaging with suppression of back-reflections from the incident beam. In order to induce an external electrostatic field in the electrooptically active substrate supporting the plasmonic coupler, the individual waveguide ends are electrically connected to a signal and ground pad, respectively. The change of the refractive index induced by the Pockels effect in z-cut LN depends on the direction of the applied electric field relative to the direction of the crystal optic axis of the crystal. To investigate the optical response on the externally applied field, the electrostatic field distribution with the field contour lines is first simulated for the two-wire configuration while one wire is grounded and the other is biased with 25V (Fig. 2). The field extends further along the depth-direction as compared to the isotropic case due to the anisotropic relative permittivity of LN. The amplitude of the electric field components along the x- and z- direction 100 nm below the LN/air interface (Fig. 2b) reveals an alternating electric field along the crystal optics axis below the individual nanostripes.

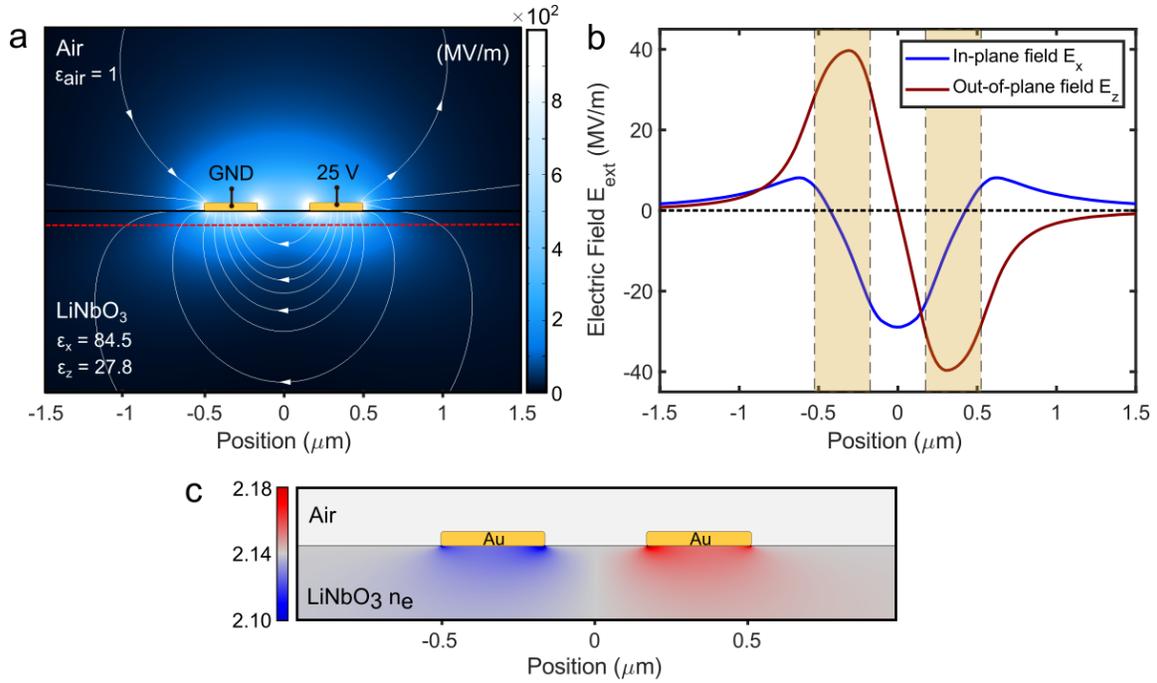

**Figure 2 | Electrostatic field simulation of the two-wire configuration. a** Field contours and electric field (color-coded) when applying a bias of 25V. The anisotropic unclamped relative permittivity of z-cut lithium niobate (LiNbO3, LN) extends the field stronger along the depth-direction, as compared to those for the isotropic case. **b** In-plane ($E_x$) and out-of-plane component ($E_z$) in the LiNbO3 substrate along the dashed red line shown in Fig. 2a. The yellow bars indicate the position of the Au wires. The in-plane field component shows highest amplitude in between the two electrodes, while the out-of-plane component exhibit highest amplitude underneath the wires with opposite sign. Due to the linear electro-optic effect in the substrate, the extraordinary refractive index underneath the left wire is decrease and increased by an equal amount in the right guide. Consequently, the waveguides lose their optical identity, which effects the power transfer in the directional coupler.

Therefore, an opposite refractive index modification is induced in the two waveguides due to the linear electro-optic effect and consequently eliminating the device symmetry of the directional coupler. An optimum overlap integral between the plasmonic field and the modulating electrostatic field is resulting in a significant phase mismatch with opposite polarities in the two waveguides, thus providing efficient push-pull operation of the directional coupler modulator. An expression for the intensity modulation and switching characteristics of the device can be derived from the coupled-mode formalism[33]. Under assumption of negligible coupling in the output bends, the normalized power in the weakly coupled waveguide system is given by

$$P_{1,2} = \frac{1}{2}\left[1 \pm \frac{2\delta L_n}{\gamma}\sin^2\left(\frac{\pi\gamma}{2}\right)\right] \quad (1)$$

with the normalized phase mismatch $\delta = \Delta\beta L/\pi$, electrically induced by the difference in the propagation constants of the individual wires $\Delta\beta = \beta_1 - \beta_2$, the interaction length $L_n = L/L_C$ normalized by the coupling length $L_C$ and $\gamma^2 = \delta^2 + L_n^2$. From equation (1), it follows that the unbiased device is inherently set to the linear section of the modulation transfer curve (quadrature point), offering high degree of linearity and collapse of even-order nonlinear distortions in the modulation spectrum. When the normalized length is set to $L_n = 1/\sqrt{2}$, the optical power can be fully switched by electrically introducing a phase mismatch of $\Delta\beta L = \sqrt{2}\pi$. Deviations from this interaction length result in reduced modulation depth and sensitivity (see Supplementary Note 1 for details). It was found that for the considered cross section geometry of the directional coupler, this condition is fulfilled at the interaction length of 15.5 µm and thus taken as the optimum design for efficient switching in our study. Due to the established material platform and the simple device geometry, the fabrication procedure of the monolithic modulator device is exceedingly simple, involving only a standard electron-beam lithography step followed by thermal gold evaporation and subsequent metal lift-off (see Methods for more details). Furthermore, the switch conceptionally exhibits relaxed tolerances to geometrical variations and wavelength instabilities compared to ring resonators or Mach-Zehnder interferometers.

**Performance of the device.** Switching and modulation of the scattering signal emitted from the output gratings of the two waveguide channels in the characterized device is illustrated with far-field images shown in Fig. 3 for two different bias voltages with opposite polarity.

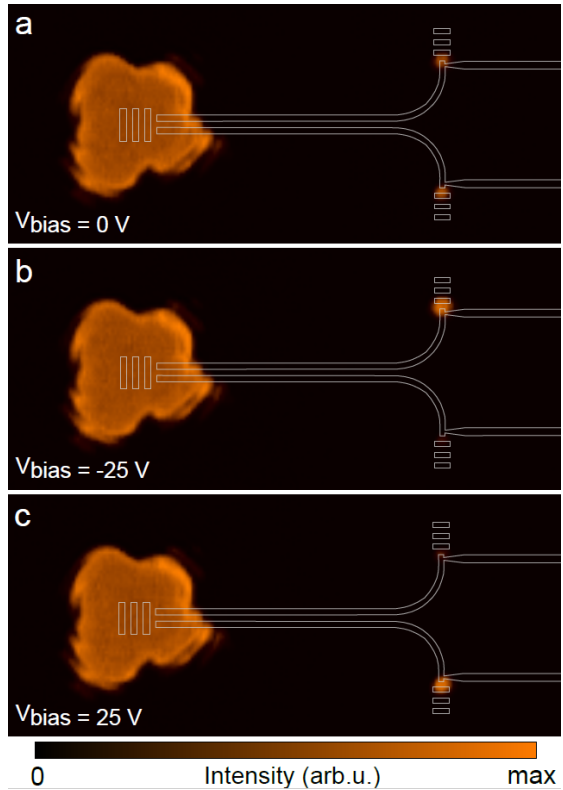

**Figure 3 | Visualization of switching and modulation behavior.** Experimental optical far-field images are captured by an infrared camera at different modulation voltages. A diffraction-limited beam (1550 nm) is positioned on the input coupling grating, to symmetrically feed the coupling system. The scattering images are superimposed with the device configuration. **a** The unbiased device exhibits equal power distribution of the scattered signal, due to optical symmetry of the directional coupler. **b** When a negative bias voltage is applied, an enhancement of the optical output signal of the biased waveguide arm is observed, while the optical signal at the opposite output port is decreased in intensity. **c** By changing the polarity of the applied voltage, the intensity in the two output terminals is switched.

At zero bias, the symmetry of the device configuration results in an equal phase condition ($\delta = 0$) in the coupler arms and consequently equal power distribution in the waveguides, revealed by a balanced scattering signal at the output ports. When a bias signal is applied, an enhancement of the output signal at one port is observed, while the signal at the opposite output port is decreased in intensity. By changing the polarity of the applied voltage, the intensity in the two output terminals is switched, revealing coupling with opposite symmetry in the directional coupler as expected from the linear modulation characteristics of the Pockels effect. The dynamic optical switching induced by an alternating voltage of $V_{AC}$ = 25 V at slow switching speeds, capturable by the infrared camera is demonstrated in Supplementary Movie 2. For obtaining the transfer curve in Fig. 4a, the applied bias is varied and the output power modulation of the two channels is measured independently at the image plane by spatial filtering with subsequent detection by a photodiode (see Methods for more details). The experimentally measured modulation curve agrees well with the theoretical transfer curve calculated by finite element method (FEM) assisted coupling mode theory (see Methods for details). A modulation depth of 90 % is achieved which corresponds to a dual-channel intensity extinction ratio (ER) of 10 dB at sufficiently high voltages, without permanent breakdown of the device. Given than full modulation depth is reached at an electrically-induced phase mismatch of $\sqrt{2}\pi$, our device exhibits a voltage-length product 0.3 V·cm, which is approximately an order of magnitude smaller than that obtained with state-of-the-art photonic LN modulators (see Supplementary Table 2 for a detailed comparison). In fact, lower levels of the voltage-length product have only been observed with electro-optic polymers, whose very low phase-transition temperatures impede their practical deployment (Supplementary Table 2). Due to non-resonant modulation characteristics and the electro-optic activity of lithium niobate which expands over a wide wavelength range, spectrally broadband operation of our directional coupler switch can be expected. The relative modulation efficiency is measured within the wavelength range of 1550 nm to 1630 nm (Fig. 4b). Although the performance of the device fluctuates with the operation wavelength, a high wavelength tolerance is observed, exhibiting less than 2 dB modulation depth variation within the investigated wavelength range.

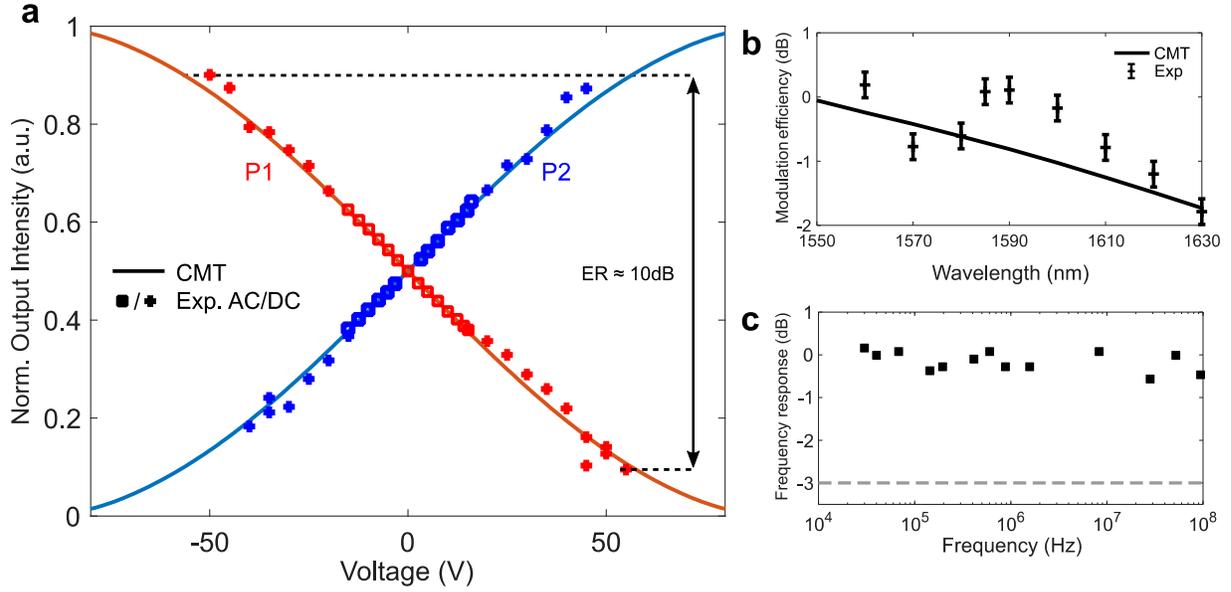

**Figure 4 | Directional coupler switch characteristics. a** Measured (symbols) and simulated (solid lines) electro-optical transfer function, showing the power exchange as a function of the applied bias voltage. **b** Wavelength dependence of the directional coupler modulator. **c** Normalized frequency response of the device.

The electro-optic frequency response of the directional coupler switch is characterized by using a high-frequency lock-in amplifier. The modulator is driven with an internal sinusoidal reference signal with a small voltage amplitude of $V_{bias} = 0.5$ V. The device exhibit a flat frequency response up to at least 100 MHz (Fig. 4c). Owing to the extremely short response time (~ fs) of the Pockels effect and the small device capacitance of only 3.6 pF, the calculated modulation cutoff frequency exceeds 800 GHz at 50 Ω resistive load, indicating potential operation at high-speeds (see Supplementary Note 3 for details).

## Discussion

In summary, we demonstrate a directional coupler switch featuring a voltage-length product of 0.3 V·cm, to date the lowest value for a device which is using lithium niobite as a material platform for electro-optic modulation. Owing to the excellent material properties such as electro-optic reliability and temperature stability, this material platform is still considered as the material-of-choice that can fulfill the demands of future optical data links, by exploring new device configurations which are significant smaller, faster and more efficient than current LN electro-optic modulators. The proposed directional coupler switch addresses this challenge with exceptional structural simplicity by providing multiport, broadband and effective modulation at compact footprint and high-speeds. The presented proof-of-concept study of utilizing integrated plasmonic circuits in LN platforms demonstrates its enormous potential, which can pave the way towards feasible communication links which hold the promise of high-speed, broadband and robust operation. The switching voltage in our configuration can potentially be further reduced by utilizing plasmonic waveguides with higher optical confinement, which can be eventually interfaced with low-loss photonic (for example, metal-diffused[4]) waveguides for practical integration into long-haul optical communication systems.

## Methods

**Device fabrication** Devices are fabricated on commercially available z-cut lithium niobite substrates. The directional coupler modulators are written by electron beam lithography using a scanning electron microscope (JOEL JSM-6490LV) at an acceleration voltage of 30 keV in 200 nm PMMA positive resist and 20 nm thick Al which serves as metallic charge dissipation layer during writing (electron doses varying between 200 and 250 µC/cm$^2$). After resist development, the directional coupler are formulated by depositing 4 nm titanium adhesion layer and 50 nm gold by thermal evaporation and subsequent 12 h lift-off in acetone. To reduce electron beam writing time, macroscopic bonding pads and connecting wires are patterned beforehand on the LN chip by optical lithography, metal deposition (5 nm Ti / 40 nm Au), and lift-off. Lithographic overlay is ensured by the mix-and-match approach.

**Experimental electro-optical characterization** A collimated and linearly polarized laser beam from a tunable IR laser (New Focus 6326 Velocity) is focused by a high numerical aperture objective (Olympus MPlan100xIR, NA = 0.95). The focused beam is positioned symmetrically on the grating coupler for symmetric feeding of the two arms of the directional coupler. The back-reflected optical signal is collected by the same objective, is passing a polarizer where the back-reflection coming from the incident beam is suppressed by aligning the polarization angle of the polarizer perpendicular to the polarization of the incident beam. This is providing an improved signal-to-noise ratio for the optical signal scattered from the output gratings which is subsequently detected by the IR camera. The integration time of the camera is adjusted to give a reasonable contrast for the output signal, leading to an oversaturation and increase in size of the incident back-reflected beam spot. By applying electrical probes on the electrode pads, one directional coupler arm is grounded while the other is biased with a AC/DC signal. For measuring the modulation transfer curve, the wavelength dependence and the frequency response of the device, the scattering signal coming from one port of the directional coupler is spatially filters and detected by a high-speed photodetector. A sketch of the experimental set-up can be found in Supplementary Note 4.

**Numerical modeling** Finite element method (FEM) simulations are performed using a commercially available software (Comsol Multiphysics 5.2a). For the electrostatic simulations, the device cross-section is modeled with the unclamped static relative permittivity tensor of LN taken from Jazbinsek et.al.[34], while the relative permittivity of air to set to be $\epsilon_{air} = 1$ and the boundaries of the two gold nanostripes are set to ground and V$_{bias}$ potential, respectively. The calculated electric field distribution is utilized to determine the modification of the refractive index in the LN substrate by using the electro-optic Pockels coefficients from Jazbinsek et.al.[34]. For the optical simulation, the modified distribution of the refractive index of LN is fed into the mode solver, while the unmodified refractive indices from Au and LN were taken from Johnson and Christy[35] and Zelmon et.al.[36]. Scattering-boundary conditions in combination with a perfectly matched layer are applied to calculate the propagation constants as a function of the applied voltage, which are used for describing the intensity modulation using the coupling-mode formalism. (See Supplementary Note 1 for details).

**Data availability**

Detailed information on the coupling mode formalism can be found in Supplementary Note 1. Detailed information on fabrication tolerances of the directional coupler switch can be found in Supplementary Note 2. The calculated device capacitance and power consumption can be found in Supplementary Note 3. Information on the experimental set-up is provided in Supplementary Note 4. An experimental near-field study of the directional coupler can be found in Supplementary Note 5 with a comparison to numerical calculations in Supplementary Table 1 and the real-space phase evolution and amplitude in Supplementary Movie 1. A comparison of state-of-the-art modulators is provided in Supplementary Table 2. Supplementary Movie 2 illustrates the dynamic optical switching at alternating voltages observed by an infrared camera.

**Acknowledgements**

The authors acknowledge funding support from the European Research Council (PLAQNAP, Grant 341054), the University of Southern Denmark (SDU2020 funding) and the VILLUM FONDEN (Grant 16498).

**Author contributions**

S.I.B. and M.T. conceived the experiment and geometry of the modulators. M.T. set up the experiment, performed the device fabrication, experimental characterization and simulations of the device. V.Z. and C.W. assisted in the device characterization and simulations. M.T. wrote the manuscript with contributions from all other authors. S.I.B. supervised the project.